\documentclass[aip,rsi,reprint]{revtex4-1}

\usepackage{amsmath,amstext}
\usepackage{multirow}
\usepackage{enumitem}
\usepackage{graphicx}
\usepackage{gensymb}
\usepackage{textcomp}
\usepackage{color}
\usepackage[normalem]{ulem}

\date{\today} 

\begin{document}

\title{DESIREE electrospray ion source test bench and setup for Collision Induced Dissociation experiments}

\author{N.~de~Ruette} 
\email{nathalie.deruette@fysik.su.se}
\author{M.~Wolf}
\author{L.~Giacomozzi}
\author{J.~D.~Alexander}
\affiliation{Department of Physics, Stockholm University, Stockholm, SE-106 91, Sweden}
\author{M.~Gatchell}
\affiliation{Department of Physics, Stockholm University, Stockholm, SE-106 91, Sweden}
\affiliation{Institut f\"{u}r Ionenphysik und Angewandte Physik, Universit\"{a}t Innsbruck, Technikerstr.\ 25, A-6020 Innsbruck, Austria
}
\author{M.~H.~Stockett}
\author{N. Haag}
\author{H.~Zettergren}
\author{H.~T.~Schmidt}
\author{H.~Cederquist}
\affiliation{Department of Physics, Stockholm University, Stockholm, SE-106 91, Sweden}

\begin{abstract}
We give a detailed description of an electrospray ion source test bench and a single-pass setup for ion fragmentation studies at the DESIREE infrastructure at Stockholm University. This arrangement allows  for collision induced dissociation experiments at center-of-mass energies between 10 eV and 1 keV. Charged fragment are analyzed with respect to their kinetic energies (masses) by means of an electrostatic energy analyzer with a wide angular acceptance and adjustable energy resolution.
\end{abstract}

\maketitle
\section{Introduction}

Interactions involving atomic, molecular, and/or cluster ions play important roles in natural plasmas \cite{Tendero2006,Yokoyama1990}, in flames and combustion processes \cite{Marr1999}, and for chemical reaction networks in interstellar media, in stellar- and planetary atmospheres, and in the comae of comets \cite{Balucani2009,Blagojevic2003}. Ion reactions are also important in biological environments \cite{Green1997,Wyttenbach2007}. Studies of these types of processes may for example help to understand how molecules and clusters are formed and destroyed in space and planetary environments \cite{Larsson2012,Tielens2013}, and how secondary ions formed by irradiation may damage neutral molecules in biological samples \cite{Prise2001,Riley1994}.

Mass spectrometry \cite{Osburn2013,Welling1998} is commonly used to study ion properties and reactions \cite{Futrell1966}. In its simplest form, a mass analyzer discriminates ions by their mass-to-charge ratio. Applying mass spectrometry in two stages, first to select the ions of interest and then selecting the products of ion-molecule reactions or dissociation processes occurring between the two stages, may reveal fragmentation schemes and compound-formation processes \cite{deHoffmann1996,KC2018}. This can help to establish the thermodynamics of specific reaction channels combined with the kinetic energy dependences of the corresponding cross sections \cite{Armentrout2001}.

Ions can be generated by various methods, depending on their properties. ElectroSpray Ionization (ESI) is an established technique to produce large molecular ions in the gas phase \cite{Fenn1964,Dole1968}. This method is very gentle and is commonly used to bring ionized complex molecules such as for examples  Polycyclic Aromatic Hydrocarbon (PAH) molecules \cite{Klaerke2013} and, especially, fragile biomolecules \cite{Fenn1989,Mann1990} into the gas phase without breaking them apart.

In the present work, we present the ESI source platform and accelerator mass spectrometer setup of the Electrospray Ionization Source Laboratory (EIS-Lab), which is part of the Double ElectoStatic Ion Ring ExpEriment (DESIREE) infrastructure \cite{Thomas2011,Schmidt2013} at Stockholm University. The EIS-Lab setup serves two purposes. It is a setup for testing different methods of producing, accumulating and cooling ions of fragile complex molecules for injection into the DESIREE ion-beam storage rings and for single-pass collision experiments with complex molecular ions. The beam-currents from ESI sources are in general very low (typically less than 1 pA for mass-selected ions) and therefore it may be necessary to accumulate them in a pre-trap before injection into a storage ring. Furthermore, compression of the ion bunch and cooling to lower internal energies of the individual ions can be performed in radio-frequency traps by means of cryogenic buffer-gas cooling. This option is available in EIS-Lab and is planned to be used on the DESIREE injection platforms. The EIS-Lab setup is also used in continuous mode to measure absolute destruction cross sections and investigate fragmentation pathways in single-pass Collision Induced Dissociation (CID) experiments. In this setup, beams of complex ions are accelerated to keV energies and made to collide with neutral atomic or molecular targets in a collision cell. The charged products are analyzed according to their energy-to-charge ratios by means of an electrostatic analyzer with large angular acceptance.

Experiments have been performed with this setup for a wide range of molecular ion projectiles, such as native and hydrogenated PAHs \cite{Chen2014, Stockett2014a, Stockett2014b, Stockett2015, Stockett2015b, Gatchell2015, Wolf2016}, fullerenes \cite{Gatchell2014,Stockett2018} and biomolecules like tetraphenylporphyrin \cite{Giacomozzi2016} and n-butylamine retinal protonated Schiff base \cite{Kostya}. Typical center-of-mass collision energies range from roughly 10~eV to about 1 keV in these experiments, which among other things have revealed the importance of fast (sub-picoseconds) single atom knockout processes \cite{Gatchell2014,Stockett2014a,Stockett2015,Giacomozzi2016,Stockett2018}. 

\section{Technical Description}

\begin{figure*}
\includegraphics[width=1\textwidth]{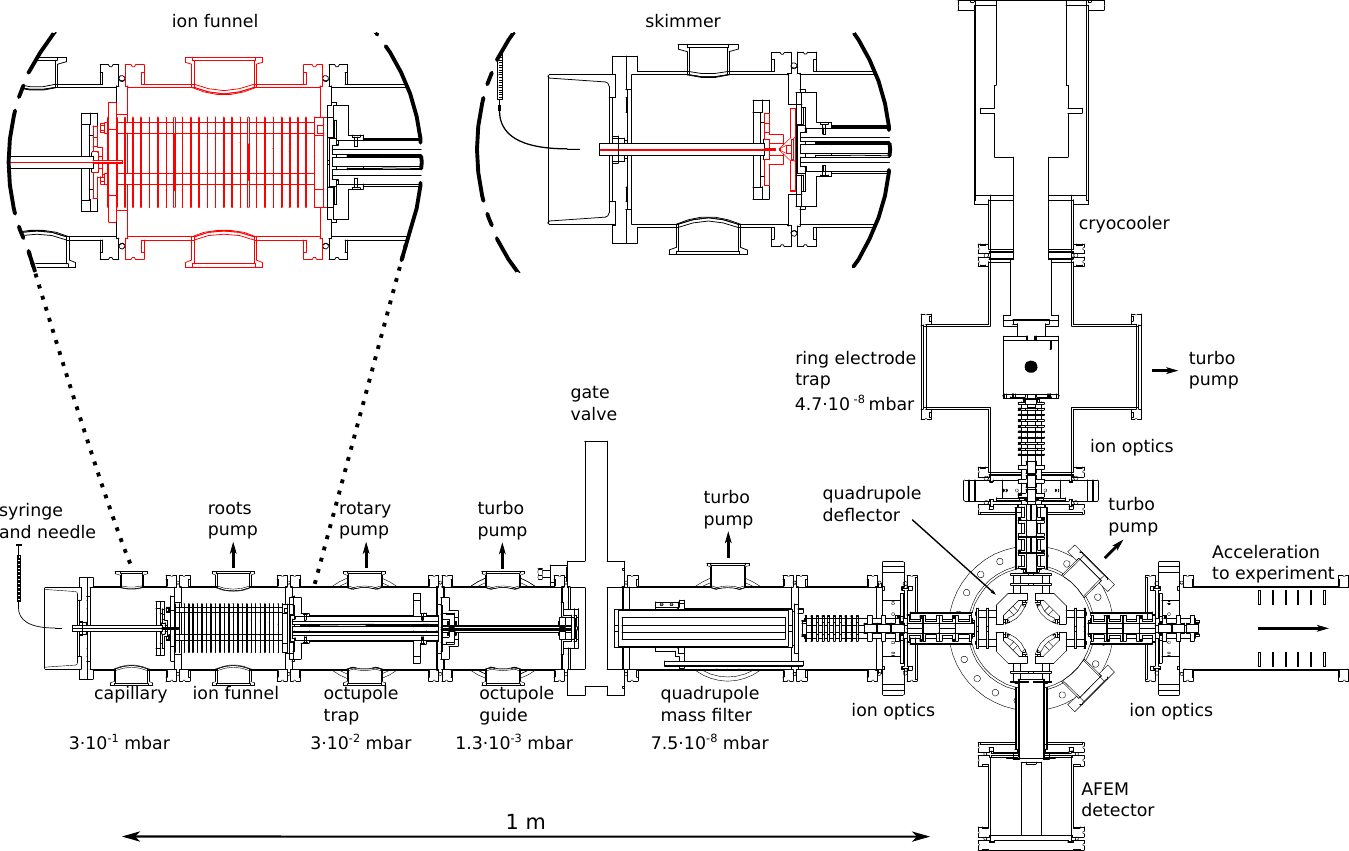} 
\caption{Schematic of the setup for ion production and mass selection. Details of the individual components are given in the supplementary material and in Table \ref{table2}. Typical pressures during operation are indicated. The insets show zoom-ins with either the ion funnel or with the skimmer in place. The choice depends on the applications (see text).} 
\label{setup1}
\end{figure*}
The setup consists of two parts: the high voltage platform for ion production, mass selection and testing of different ion-beam production methods, and the beamline for CID experiments.

In the first part, operated on a 20 kV high voltage platform, molecular ions are produced by the ESI source, and subsequently pass through an ion funnel, octupole guides, a quadrupole mass filter and a quadrupole deflector with three of its four ports equipped with horizontal and vertical beam deflectors and lenses. The different stages in the ion production and preparation beam-line are separated by small apertures for differential pumping. An Active Film Electron Multiplier (AFEM) detector, mounted below the quadrupole deflector, can be used to monitor the ion production and record spectra of the mass-selected ions reaching this position. Detailed descriptions of the various components of this first part are given in section~\ref{platform}. 

In the second part, situated after the high voltage platform, the accelerated and mass-to-charge selected molecular ion beam enters the experimental section. This section contains optics for steering and focusing, and a collision gas cell. Two Einzel lenses and two pairs of electrostatic deflector plates serve as a large angular acceptance energy analyzer. The products are then directed to a position-sensitive microchannelplate detector. The kinetic-energy-per-charge fragment spectrum is obtained by combining the measured position on the detector and the energy analyzer voltage settings for each individual fragmentation event. As the relative change of the projectile velocity is usually small, this measured spectrum is equivalent to the mass-to-charge distribution of the product ions.

Positive or negative ion beams can be produced by choosing the appropriate sample solution for the electrospray and by setting the correct polarities and voltages on the guiding and focusing elements. All the potentials on the high voltage platform and the following beamline can easily be reversed to obtain a beam of the same species but with the opposite charge state. This only requires minor adjustments of the absolute values of the voltage settings. The polarity of the analyzed fragments can also be selected independently of the platform settings, \emph{e.g.} positive fragments from CID of negative primary ions can be analyzed.

\subsection{Ions production and mass selection} \label{platform}

A schematic of the setup on the high voltage platform, including the acceleration stage, is shown in Fig.~\ref{setup1}. Details on commercial components, including the vacuum equipment, are given in the supplementary material. Specifications of custom-built components are given in Table~\ref{table2} and schematics are shown in the subsections below.

\newcommand*{\TableIndent}{\hspace*{0.5cm}}
\begin{table}
\caption{Custom-built components mounted on the high voltage platform.}
\label{table2}
\begin{tabular}{|l|c|}
\hline 
\multicolumn{1}{|c|}{Item} & Notes \\ 
\hline
\hline 
 \multicolumn{2}{|c|}{\underline{Ion funnel}} \\
Electrodes & 26 \\
\TableIndent Material & Stainless steel \\
\TableIndent Thickness & 0.5 mm \\ 
\TableIndent Inner diameter & $11\times 38$ mm, then decreasing \\ 
 & in 2 mm steps from 36-8 mm \\
\TableIndent Spacing & 5 mm \\ 
Coupling Capacitors & 1 nF \\
Resistors & 10 M$\Omega$ \\ 
\hline
\multicolumn{2}{|c|}{\underline{Octupole Trap}$^*$} \\
\TableIndent Length & 180~mm \\
\TableIndent Inner diameter & 16.9~mm \\ 
Rods & \\
\TableIndent Material & Stainless steel \\
\TableIndent Diameter & 6 mm \\
\hline
\multicolumn{2}{|c|}{\underline{Octupole Guide}} \\
\TableIndent Length & 158.5~mm \\
\TableIndent Inner diameter & 5.6~mm \\ 
Rods & \\
\TableIndent Material & Stainless steel \\
\TableIndent Diameter & 2 mm \\
\hline
\multicolumn{2}{|c|}{\underline{Ion optics}} \\
Acceleration electrodes & 10 \\
\TableIndent Material & Stainless steel \\
\TableIndent Thickness & 2 mm \\ 
\TableIndent Inner diameter & 15 mm \\
\TableIndent Spacing & 5 mm \\ 
\TableIndent Resistors & 1 M$\Omega$ \\ 
Einzel Lens & \\ 
\TableIndent Material & Stainless steel \\
\TableIndent Thickness & 5 mm \\ 
\TableIndent Inner diameter & 10 mm \\
\TableIndent Spacing & 2 mm \\ 
Deflectors & \\ 
\TableIndent Material & Stainless steel \\
\TableIndent Length & 19 mm \\ 
\TableIndent Inner diameter & 12 mm \\
\TableIndent Pair distance & 1 mm\\
\TableIndent Spacing & 2 mm \\ 
\hline
\multicolumn{2}{|c|}{\underline{Ring Electrode Trap}} \\
Electrodes & 30 \\
\TableIndent Material & Gold covered copper \\
\TableIndent Thickness & 1 mm \\ 
\TableIndent Inner diameter & 20 mm \\
\TableIndent Spacing & 1 mm \\ 
Coupling Capacitors & 1 nF \\
Resistors & 3 M$\Omega$ \\ 
\hline 
\end{tabular}\\
\vspace{0.2cm}
{$^*$ Ref. \cite{Taban2005} gives a detailed description of the octupole trap.}
\end{table}

\subsubsection*{The ESI source}

The ESI source consists of a stainless steel needle mounted in air followed by a heated stainless steel capillary as an interface to the vacuum region. 

A motor-driven syringe provides a constant flow of a sample solution, typically around 1--2 {\textmu}l/min to the needle via a fused silica wire. Recipes (solvent, concentration of the molecule of interest, ionizing agent, etc.) vary depending on the analyte (the chemical substance to be studied) and can usually be found in the literature. As an example, to produce protonated adenine, a nucleobase which is easy to spray and often used for tests, the sample is dissolved in a mixture of water and methanol in a 50:50 volume ratio with the addition of 5\% acetic acid \cite{MAUTJANA2009}. A typical concentration of Adenine in the solution is 0.1 mmol/l. For PAH cations, we typically mix 100~{\textmu}l of a 0.5~mmol/l solution of the sample dissolved in dichloromethane with 500~{\textmu}l of a 0.1 mmol/l solution of silver nitrate -- used as an ionizing agent -- in methanol \cite{marziarz05}. For anion production, the acid (proton donor) or ionizing agent is usually replaced by a base (proton acceptor) or by an electron donor. For example, we use tetrathiafulvalene dissolved in dichloromethane to produce \cite{tomita02} C$_{60}^-$ and ammonium hydroxide (with acetonitrile replacing the methanol) to produce de-protonated adenine \cite{Wincel2016}. 

The needle is set on a potential of typically 3-4 kV with the appropriate polarity (as for all the potentials mentioned hereafter), while the capillary is biased at 100-300~V. The ions enter a low-vacuum region through this capillary which has an inner diameter of 0.48 mm and length of 130 mm. A pressure of $3 \times 10^{-1}$ mbar is maintained with a 1000~m$^3$/h Roots pump in this first pumping stage. The capillary consists of a stainless steel rod, which is tightly fitted inside a copper tube for thermal conductivity. The copper tube is usually heated to a temperature of around 120$^{\circ}$C, however this temperature can be lower in some cases (e.g. for the production of solvated ions). The capillary is inserted in an electrode placed towards the exit to accelerate the ions by the application of a higher/lower potential (up to 150 V more/less) than the capillary itself, for positive and negative ions respectively. This increases the ion transmission to the next section.

\subsubsection*{Ion funnel or skimmer setup}

\begin{figure} [h!t]
\includegraphics[width=.45\textwidth]{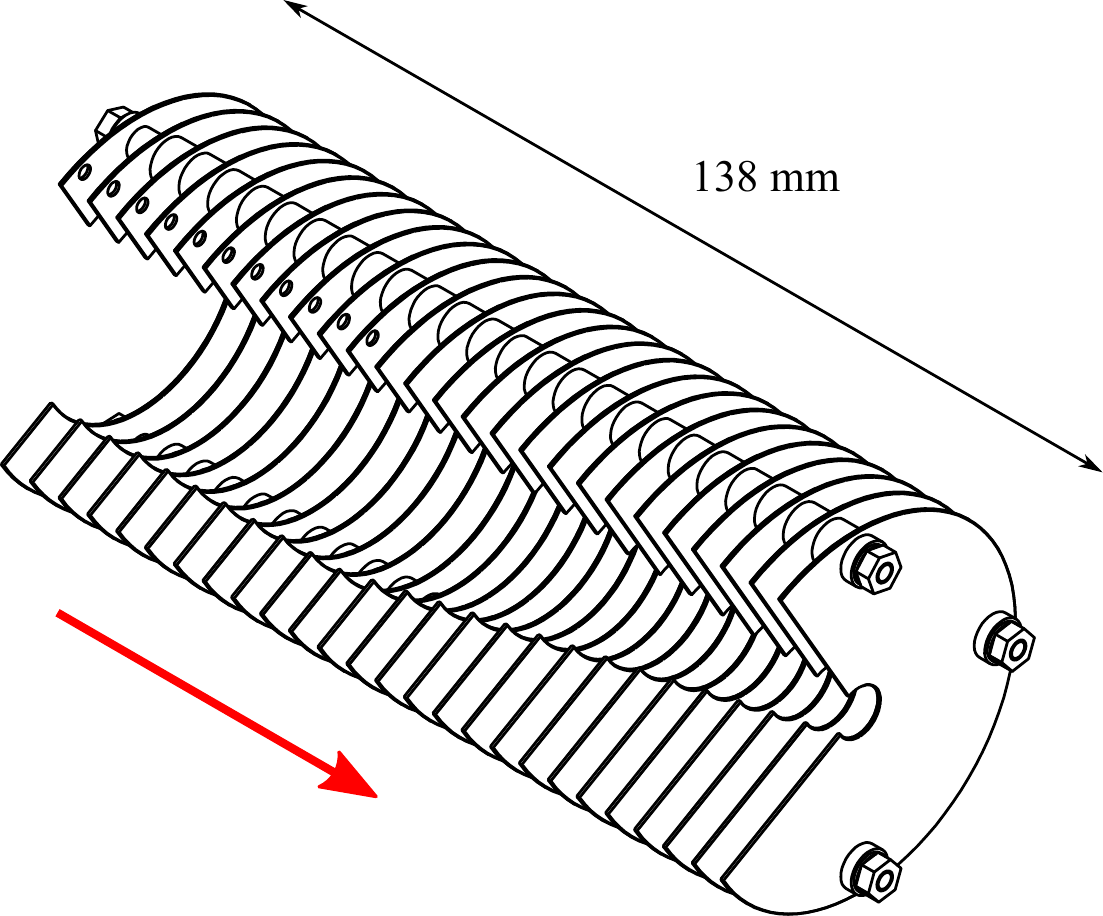}
\caption{Schematic cut-away drawing of the ion funnel. Dimensions are given in Table \ref{table2}. The red arrow indicates the ion propagation direction.} 
\label{funnel} 
\end{figure}

\begin{figure} [h!t]
\includegraphics[width=.35\textwidth, trim = 1cm 1cm 1cm 1cm]{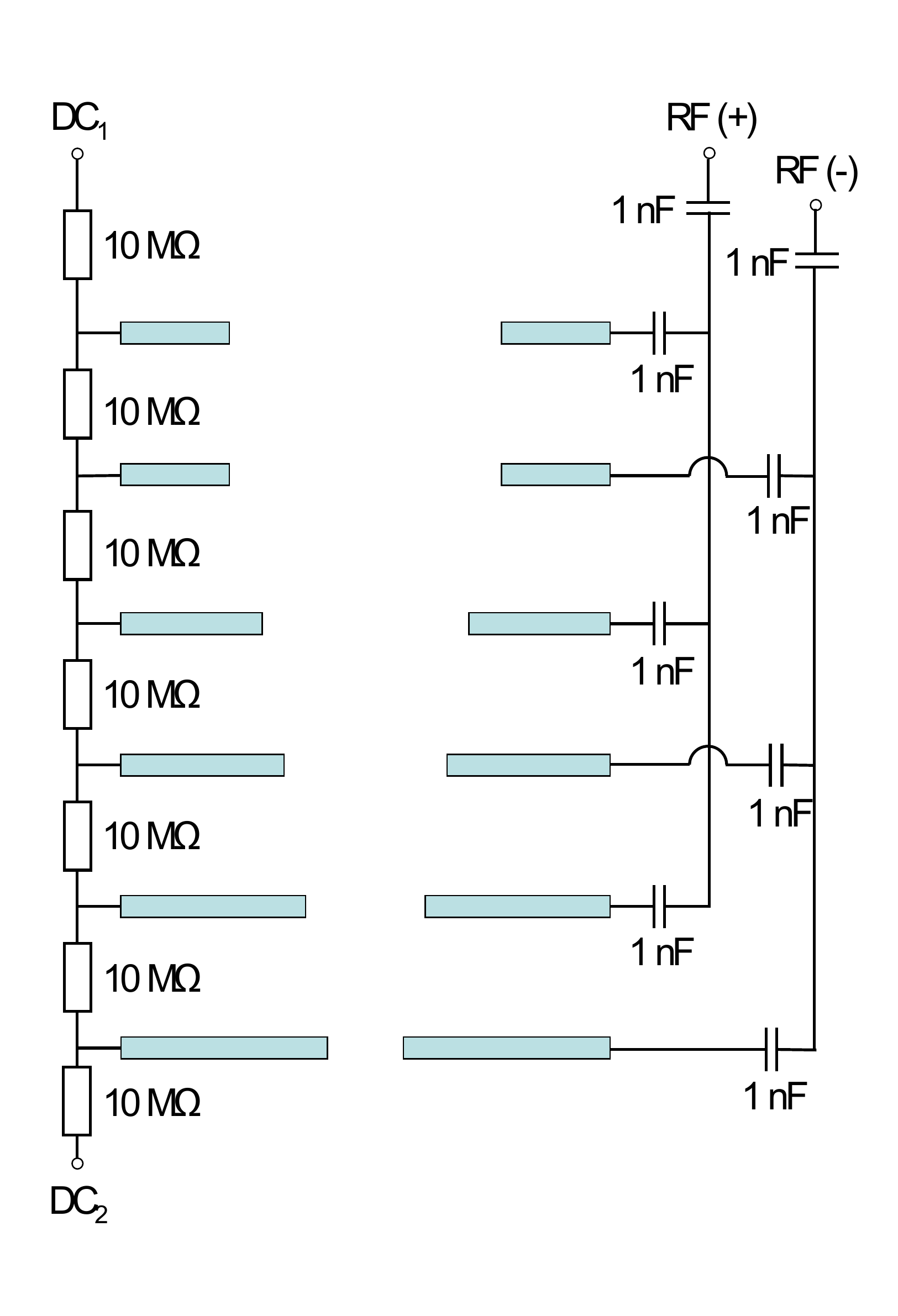}
\caption{Schematic drawing of the RF and DC circuits for the ion funnel.} 
\label{circuit} 
\end{figure}

In order to collect the ions after the capillary, the ion source is followed by an electrodynamic ion funnel \cite{Shaffer1997,Shaffer1998,Shaffer1999} based on the design of Julian \emph{et al.} \cite{Julian2005}. The 26 ring electrodes forming the funnel (see Fig. \ref{funnel} and Table \ref{table2} for details) are connected by 1 nF coupling capacitors and 10 M$\Omega$ resistors mounted under vacuum (see the schematic in Fig. \ref{circuit}). The ion funnel provides a collisional focusing effect towards the beam axis and thus increases the ion transmission through the application of a radio frequency (RF) and DC fields with a background gas pressure of the order of $10^{-1}$~mbar \cite{Douglas1992}. The 1.75 MHz RF potential of variable amplitude is applied with opposite phase on adjacent electrodes creating a time-independent effective potential $V^*$ that radially confines the ions (see the review by Gerlich \cite{Gerlich2007}). At the same time, the DC potential gradient, created by the resistor chain, along the funnel axis drives the ions toward the exit. The DC voltages are kept close to the bias on the capillary and capillary end-plate. The specific feature of the Julian \emph{et al.} funnel design \cite{Julian2005} is the larger spacing between the ring electrodes compared to most other such devices. We chose this solution \cite{Julian2005} for easy manufacturing and assembly and for its enhanced ion focusing properties.
 
In our setup, we also have the option to use a skimmer (0.75 mm diameter opening) instead of the funnel, as shown in the insets in Fig. \ref{setup1}. However the funnel offers a higher collection efficiency (about a factor 3). Furthermore, collisions with the residual gas in the funnel region can help to evaporate solvent molecules and to break unwanted clusters. The skimmer may, on the other hand, be useful for studies of solvated molecular ions and clusters. This was the case for example with clusters from the ProteoMass LTQ/FT - Hybrid ESI Positive Mode Calibration Mix with a mass range from m/z 138 to m/z 1822. For these applications, we remove the chamber containing the funnel, replace the subsequent electrode (entrance of the octupole trap presented below) by the skimmer and use a shorter, 110 mm long capillary, as shown in the insets in Fig. \ref{setup1}.

\subsubsection*{Ion guiding, accumulation and mass selection} \label{guiding}

\begin{figure} [h!]\includegraphics[width=.47\textwidth]{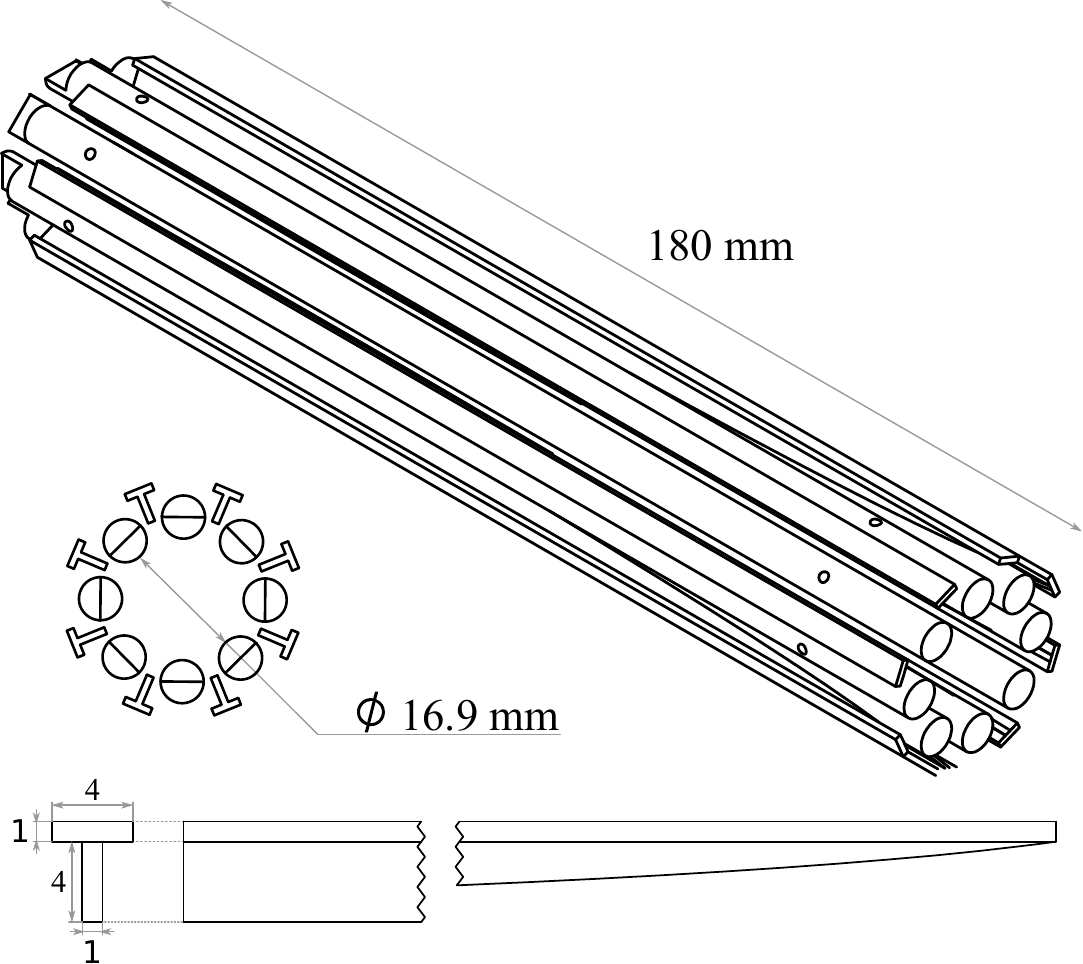}
\caption{A schematic of the octupole trap is shown on top of the figure (without end electrodes) and a cross section of the entrance is in middle left. An ejection electrode's profile (front and side view) is shown in the lower part of the figure. Dimensions are given in table \ref{table2}.} 
\label{octtrap} 
\end{figure}

In the stage following the ion funnel, a linear RF octupole (see Table \ref{table2} and Fig. \ref{octtrap}) can be used either as an ion guide (continuous operation) or as an ion trap (pulsed operation) in order to form ion bunches. Continuous and pulsed operation are used respectively for CID experiments and for accumulation of ions for injection into the DESIREE-storage rings or into various ion-beam traps in EIS-Lab. 

The octupole was custom-built after a design by Taban {\em et al}. \cite{Taban2005}. Confinement for the pulsed-operation mode is achieved by a combination of an RF potential (1~MHz, 200 V$_\text{pp}$) applied on the rods such that each rod has a potential 180$^\circ$ out-of-phase with its neighbor, and a DC potential on electrodes placed at the entrance and exit ($50-80$ V on the entrance and a lower voltage on the exit give the maximum ion transmission depending on the type of ions). A DC pole bias (similar to the entrance voltage) is also applied on the rods. The ions' kinetic energies are initially sufficient to overcome the potential barrier at the entrance of the trap. Collisions with the residual gas then reduce their kinetic energy, effectively trapping the ions. Then the barrier at the exit has to be lowered to release them from the trap. The extraction of short pulses of ions (about 200 {\textmu}s) is facilitated by ejection electrodes specifically designed by Taban {\em et al}. \cite{Taban2005} with T-shaped cross sections and a curved profile in the axial direction as shown in Fig.~\ref{octtrap}. They are mounted in between the rods and push the trapped ions toward the exit by creating a small potential gradient.

Differential pumping between the funnel and octupole chambers is achieved with a 2.0 mm diameter aperture in the entrance DC electrode of the octupole. A rotary vane pump is used to obtain a pressure of $3 \times 10^{-2}$~mbar in this section (see Fig. \ref{setup1}).

\begin{figure} [h!t]
\includegraphics[width=.45\textwidth]{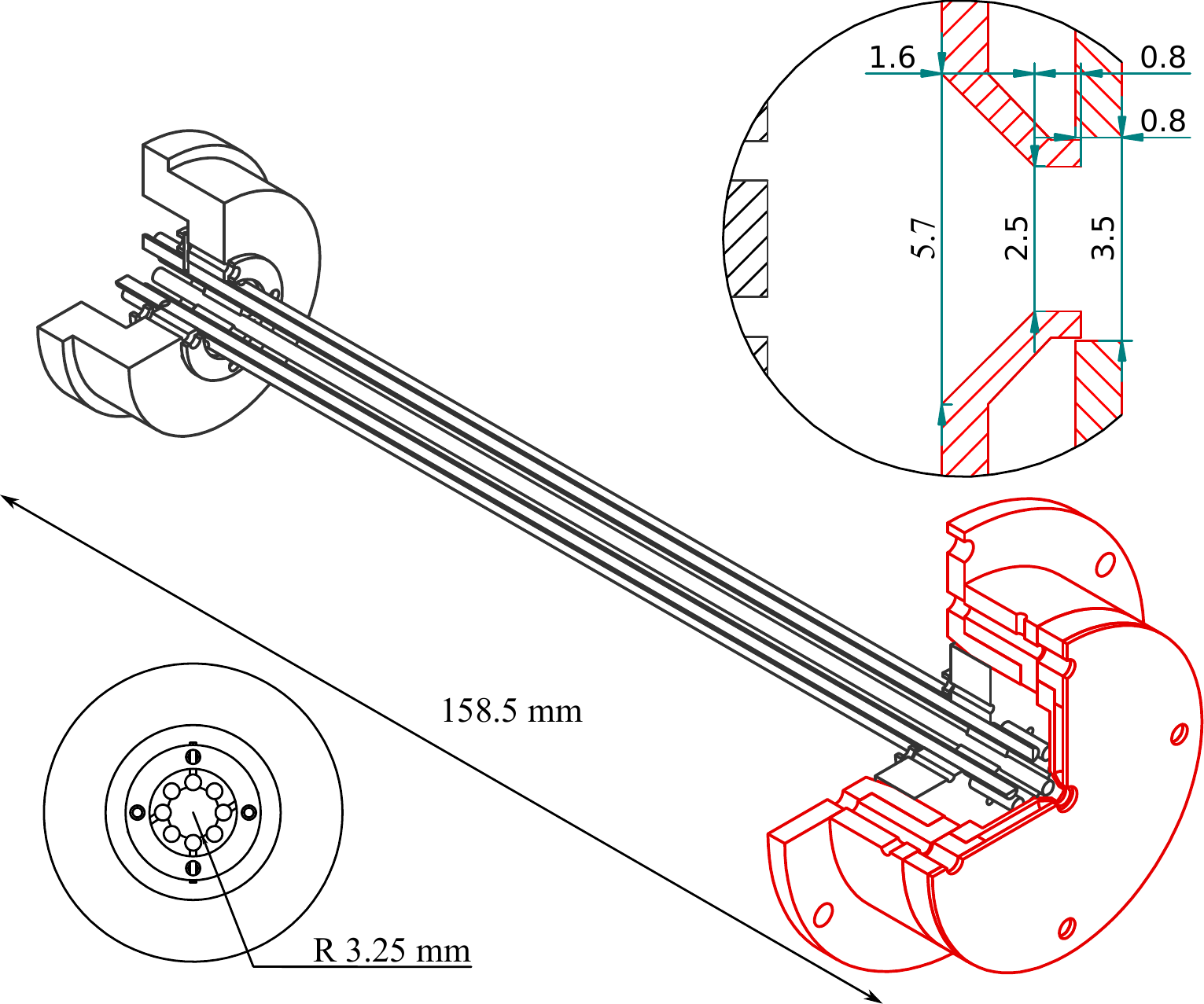}
\caption{Schematic of the octupole ion guide with lens in red. The dimensions of the cone on the lens are shown in the inset to the upper right. A cross section of the entrance is shown in the bottom left of the figure.} 
\label{octguide} 
\end{figure}

At the exit of the octupole guide/trap, the ions pass through a 1.5 mm aperture and are guided through an additional differential pumping region ($1.3 \times 10^{-3}$ mbar) by a second, 158.5~mm long, octupole guide (see Fig. \ref{octguide} and Table~\ref{table2}). The RF-voltage applied to the rods has a frequency of 2.4~MHz and an amplitude of 200~V$_\text{pp}$. A small cone-shaped electrode with an aperture of 2.5~mm in diameter and a planar ring electrode are mounted after the octupole guide and act as a lens (see Fig. \ref{octguide}). It increases the ion transmission through a gate valve placed behind, which introduces a gap of 44 mm without any possibility to guide the ions. Another way to improve the transmission of slow ions would be to minimize the gate valve thickness and to electrically insulate it to allow the application of a guiding potential \cite{Pittman2005}. The valve is used to isolate the source region in order to clean the needle and capillary with a high flux of pure methanol (150 {\textmu}l/min) on a daily basis and thus prevent their clogging. The gate valve also makes it possible to vent the source region without breaking the downstream vacuum ($7.5 \times 10^{-8}$~mbar).

After the gate valve, the ions are mass selected by an Extrel CMS quadrupole mass filter (see the supplementary material) with a m/z range of 2-4000 amu/e. The combination of RF and DC voltages determines the stable ion trajectories through the filter for each mass \cite{March1997}. As the resolution is increased, the transmission through the quadrupole decreases due to various factors including the ion energy, the ion optics, which control the distance of an ion trajectory from the quadrupole axis and their transverse energy, and the accuracy of the machining and mounting of the octupole rods. A compromise between high mass resolution and high transmission needs to be reached and will depend on the experiment. For example in the study of hexahydro-pyrene cations (C$_{16}$H$_{16}^+$, m/z~=~208~amu/e) \cite{Gatchell2015,Wolf2016}, a good mass resolution was needed in order to resolve the loss of a single H-atom. 

\begin{figure} [h!t]
\includegraphics[width=.45\textwidth]{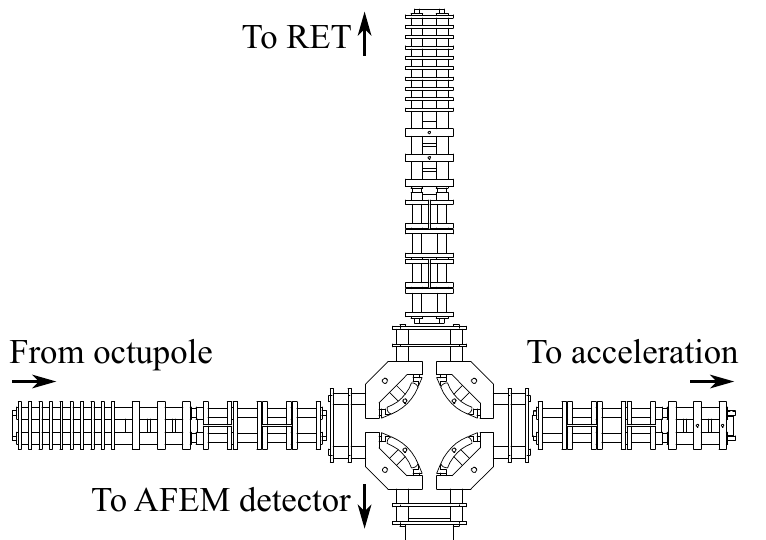}
\caption{Schematic of the quadrupole deflector with ion optics.} 
\label{quad-defl}
\end{figure}

\begin{figure} [h!t]
\includegraphics[width=.45\textwidth]{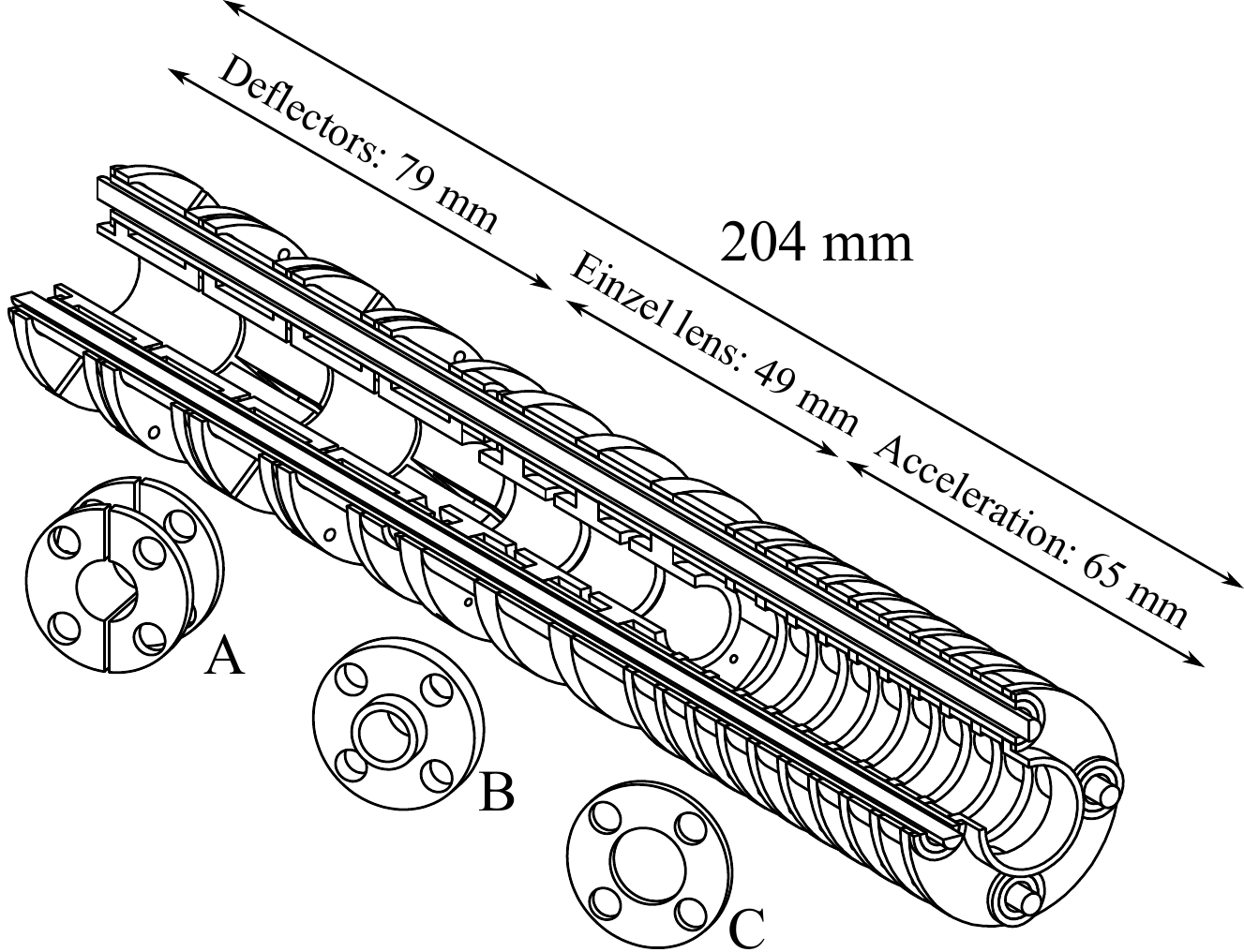}
\caption{Schematic of the ion optics mounted on the quadrupole deflector. A: Single deflector unit, B: Single lens unit, C: Single acceleration disk. Dimensions are given in table \ref{table2}.} 
\label{ion_optic} 
\end{figure}

After the mass filter, a quadrupole deflector (see Fig~\ref{quad-defl}) directs the ions either onto an AFEM detector to measure the intensity of the mass-selected ions, a cryogenic Ring Electrode Trap (RET) to compress the ion bunches and to cool the ions internally and translationally (see below), or the acceleration stage that directs the ions towards the CID experimental section. At the entrance of the quadrupole deflector, the ions are accelerated to a kinetic energy of 200 eV per atomic unit of charge by 10 equidistant electrodes connected by a resistor chain ($9 \times 1$ M$\Omega$) to provide a homogeneous acceleration field. An Einzel lens and two sets of horizontal and vertical beam deflectors together with the acceleration unit form the ion optics assembly shown in Fig. \ref{ion_optic} (see  also Table~\ref{table2} for details). The whole assembly is mounted on two of the four entrances/exits of the quadrupole deflector and an assembly with only the lens and deflectors is mounted in the forward direction (see Fig \ref{quad-defl}). 

The AFEM is used to monitor the rate of mass-to-charge selected ions that are transported through the quadrupole bend. This discrete-dynode type detector was chosen for its linear response even at high count rates up to $5 \times 10^6$ Hz. For negative ion detection, a positive voltage is applied on the first dynode to ensure that slow ions reach the detector with sufficient energy to be detected. 

When leaving the high voltage platform, the ions are accelerated towards the ground-potential ion beam-line. The ion source platform is isolated from ground and can be set to a maximum potential of $\pm$20 kV. Electronics, pumps and other electrical devices placed on the platform are powered through a 20 kV isolation transformer. Data communication to/from the high-voltage platform is achieved through fiber optics.

\subsubsection*{Ring Electrode Trap} \label{trap}

\begin{figure} [h!t]
\includegraphics[width=.45\textwidth]{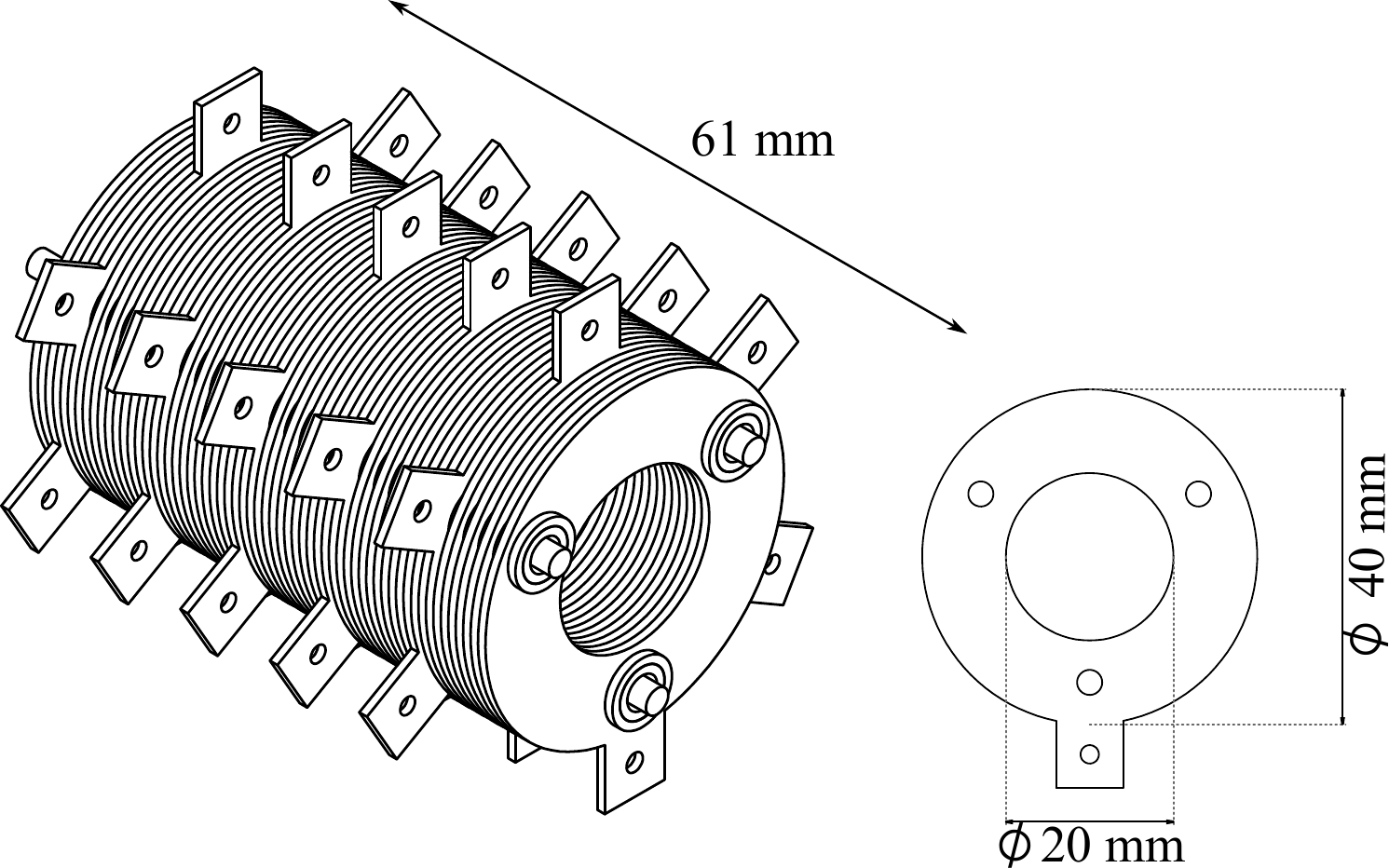}
\caption{Schematic of the ring electrode trap and one of the 30 identical electrodes. Dimensions are given in table \ref{table2}.} 
\label{RET} 
\end{figure}

\begin{figure*}[h!t] 
\includegraphics[width=1\textwidth]{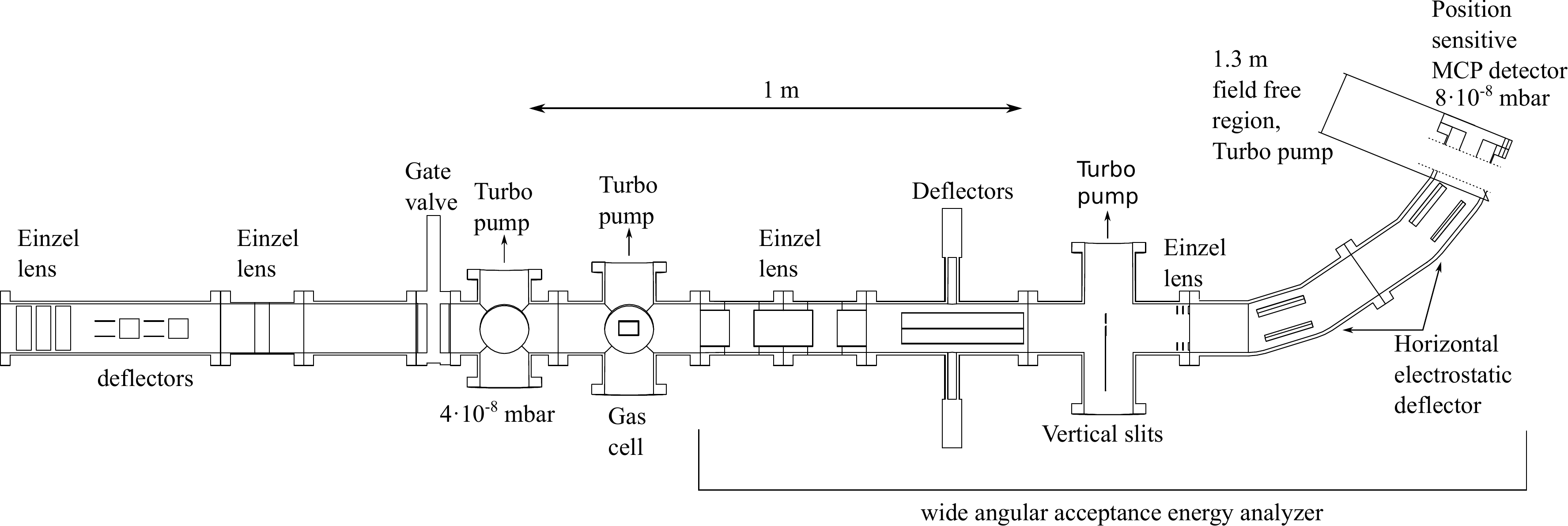}
\caption{Schematic top view of the beamline for experiments on ground potential.}
\label{setup2}
\end{figure*}

In order to inject sufficient numbers of electrospray produced ions into the DESIREE storage rings \cite{Thomas2011,Schmidt2013}, the ion-accumulation and pulsed-mode operation has to be optimized in terms of ion-pulse length and the number of ions. The shortest bunch that can be extracted from the RF-octupole trap is about 200 {\textmu}s which is too long for the rings in DESIREE. As an example, a pyrene (202~amu) beam at 10~keV would take approximately 90 {\textmu}s to fill up one of the rings (8.6~m in circumference) \cite{Thomas2011}.
 
To compress the bunch further, both spatially and temporally (down to a 15 {\textmu}s pulse length), and to cool the ions internally \cite{Otto2013}, a cryogenic Ring Electrode Trap (RET) \cite{Luca2001} has been mounted at one of the exits of the 4-way quadrupole deflector (see Fig. \ref{setup1} and Fig. \ref{RET}). After deceleration (see Fig. \ref{ion_optic} and Fig. \ref{quad-defl}), the ions can be trapped by an RF potential (1~MHz, 200 V$_\text{pp}$) in combination with a DC potential at the end electrodes. Similar to the ion funnel, the 30 ring electrodes are electrically insulated and connected by 3~M$\Omega$ resistors (DC voltages) and 1~nF coupling capacitors (for the RF signals) situated, in this case, outside the vacuum chamber on a separate electrical circuit board. 

The RET is mounted inside a copper housing connected to the cold head of a cryogenerator, and can be cooled down to 5~Kelvin. Helium buffer gas (0.12 l) is admitted prior to ion injection through a pulsed poppet valve and quickly reaches thermal equilibrium with the trap. Collisions with the cold He atoms dissipate the kinetic energy of the injected ions and additionally bring the ions' internal degrees of freedom into thermal equilibrium with the gas. Before extraction of the cold molecular ions, the buffer gas is pumped away through the 10 mm diameter entrance/exit aperture of the trap  (from 0.1 mbar to 10$^{-3}$ mbar inside the trap) in about 200 ms. This is to prevent re-heating of the ions by energetic collisions with He during the acceleration phase. 

The trap design was simulated with CPO (Charged Particle Optics software from Electron Optics, https://electronoptics.com/). This program is capable of handling space charge effects with static fields. The simulations showed that it should be possible to extract ion bunches with a pulse width of about 15 {\textmu}s and an energy spread of about 10 eV \cite{Haag2011}. Experimental tests of the pulsed-mode operation have not been fully performed yet.

\subsection{Beamline for CID experiments}

The beamline and setups for wide angular acceptance CID experiments are shown schematically in Fig.~\ref{setup2}. Specifications of the main components are listed in the supplementary material. 

Ion optics consisting of an Einzel lens, four sets of parallel-plate electrostatic deflectors (two horizontal, two vertical), and a second Einzel lens are used to steer and focus the ion beam for transport through a $4.0 \pm 0.2$ cm long gas cell with entrance and exit apertures of 1.0~mm and 1.5~mm in diameter, respectively. The pressure in the gas cell is regulated by a manual needle valve and measured using a gas independent ceramic capacitance gauge. Three turbo pumps, one before, one at, and one after the gas cell chamber, help to limit the rise in background pressure as the target gas is introduced. 

Most targets that are in gaseous form at room temperature can be used and it is also possible to evaporate liquids into the gas line for some experiments. With beam energies ranging from 700 eV to 15 keV in the laboratory frame of reference, center-of-mass energies from keV levels down to a few tens of eV may be reached. For example, 0.7-15~keV pyrene molecular cations (C$_{16}$H$_{10}^+$, mass~=~202~amu) collide at center-of-mass energies of 13-290 eV and of 63-1350 eV with He and Ne targets, respectively.

Charged collision products (fragments) and intact ions exit the gas cell with nearly the same velocity and are analyzed according to their energy-to-charge ratios by means of a wide angular acceptance energy analyzer. This analyzer consists of two Einzel lenses to focus the ions, a set of vertical and horizontal deflectors for small corrections of the beam direction and a pair of large angle horizontal electrostatic deflectors. A set of narrow vertical slits placed in front of the second Einzel lens and the horizontal deflectors (see Figure \ref{setup2}) can be used to improve the resolution at the expense of the angular acceptance. After the deflectors, the ions are detected on a 40~mm double-stack MicroChannel Plate (MCP) with a position sensitive resistive anode. The kinetic-energy-per-charge spectrum is obtained from the list mode recording of the position on the MCP for each individual hit and the voltage settings of the horizontal deflectors at this moment in time. This is described in more detail in Section \ref{CID-spectrum}.

\section{Experimental procedure and tests}

\subsection{Ion funnel tests} \label{funneltest}

Tests have been performed on the ion funnel to optimize the throughput as functions of the RF and DC voltages and the background gas pressure \cite{Haag2011} using the AFEM detector.

The first tests described hereafter concern the RF voltage. The effective potential $V^*$ confining the ions in the guide is proportional to $1/ (mf^2)$ where $f$ is the RF frequency and $m$ the mass of the ions. This effective potential also depends on the distance $d$ between the electrodes relative to the inner diameter of the exit plate $\rho$ (see Eq. (2) in Kelly {\em et al.} \cite{Kelly2010}). Previous studies on ion funnel design \cite{Lynn2000,Tolmachev2000} showed that ions can get trapped in the effective potential well between the plates. The spacing between the rings and the thickness of the electrodes must be significantly smaller than their inner diameter to reduce the depth of this effective potential well to avoid trapping. The value of these parameters ($f, m, d, \rho$) can change the transmission and the mass window of the ions dramatically \cite{Kelly2010}.

The sensitivity to the RF frequency was tested using two different power supplies, one at $f=$~915~kHz and the other at 1.75 MHz. The funnel was found to offer a better ion transmission at the higher frequency, which was then used for the tests described below.

\begin{figure}[h!t]
\includegraphics[width=.47\textwidth, trim=1cm 2cm 1cm 2cm]{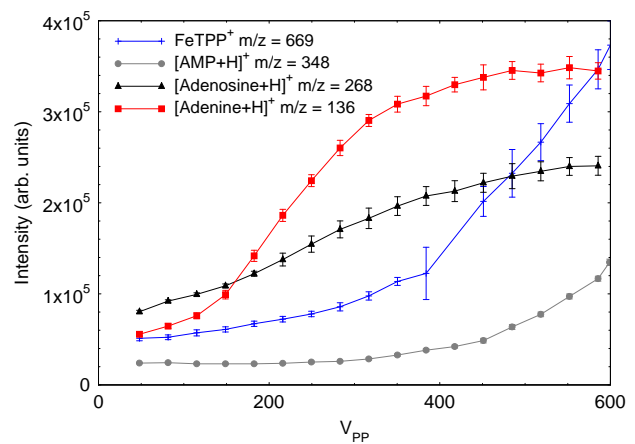} 
\caption{Ion beam intensities as a function of the ion-funnel RF voltage peak-to-peak amplitude, with an RF frequency of 1.75~Mhz, for different biomolecules: protonated-adenine (m/z = 136, red squares), protonated-adenosine (m/z = 268, black triangles), protonated-adenosine-monophosphate ([AMP+H]$^+$, m/z = 348, blue cross) and 5,10,15,20-tetraphenyl-21H,23H-porphinato iron(III) cations (FeTPP$^+$, m/z = 669, gray circles). The lines between data points are to guide the eye. The statistical errors are one standard deviation.}
\label{Vppmass}
\end{figure}

Beam intensities as a function of the RF amplitude were measured for ions of different masses (from m/z~=~82 to 669). Each beam was optimized for maximum transmission to the AFEM before scanning the peak-to-peak amplitude V$_{PP}$ of the RF voltage. Fig.~\ref{Vppmass} shows the beam intensity of four different beams of biomolecules: the protonated-nucleobase adenine (m/z = 136), protonated-adenosine (m/z = 268), protonated-adenosine-monophosphate ([AMP+H]$^+$, m/z~=~348), and 5,10,15,20-tetraphenyl-21H,23H-porphinato iron(III) cations (FeTPP$^+$, m/z~=~669) as a function of V$_{PP}$, with an RF frequency of 1.75~MHz. In all four cases, the intensity increases with the RF voltage. For the lower mass ions, the signal reaches a plateau. However it is not observed for the heavier ions ($\gtrsim$~200~amu/e) for this range of voltages. Moreover, the voltage where the intensity of the AMP and FeTPP begins to increase is higher than that required for adenine and adenosine. High masses seem to require a higher voltage than what can be safely applied to the funnel to reach optimum transmission. We did not observe ions with masses above 900 amu/e although they were detected when the funnel was replaced with the skimmer. This could be explained by a limited transmission window due to the limited voltages range of the RF power supply together with the harsh conditions in the funnel, leading to collisional destruction of loosely bound ions. However, our funnel does not have the limitations of other type of funnels with smaller inner diameters which limit the transmission of ions with less than a few hundred amu \cite{Shaffer1997,Shaffer1998}. This is due to the depth of the axial effective potential well as it varies as $1/m$ and increases toward the funnel exit as the ring electrode apertures become comparable to the spacing between the electrodes \cite{Shaffer1999,Lynn2000,Tolmachev2000,Kelly2010} (see Table \ref{table2}).

Applying a DC voltage, in addition to the RF voltage, to the funnel is known to offer better control on the transmission than gas dynamic effects alone \cite{Shaffer1999}. Lowering the axial DC field improves the transmission of light ions \cite{Page2006}. Changing the settings of the DC voltages at the entrance and exit of the funnel changes the conditions and can result in internal heating or even fragmentation of molecular ions \cite{Julian2005}. This was for example observed with hydrogenated pyrene where the loss of hydrogen atoms could be slightly reduced by lowering the potential on the electrodes (around 100 V) and on the capillary. We have also noticed that the DC gradient on the funnel axis does not need to be large and the optimized values do not often differ by more than 10~V across the funnel length. 
 
\begin{figure}[h!t]
\includegraphics[width=.49\textwidth, trim = 1.5cm 1cm 0cm 2cm]{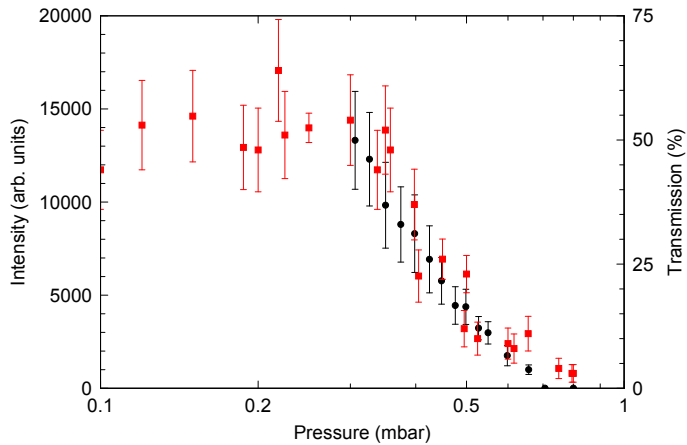}
\caption{Measurements of the beam intensity for pyrene cations (C$_{16}$H$_{10}^+$, mass~=~202~amu) as a function of the background gas pressure in the ion funnel, shown by the black circles and left Y-axis. The red squares show the simulations of the transmission probabilities (right Y-axis) of the same ions, through the funnel. The statistical errors are one standard deviation.}
\label{funnel-pressure}
\end{figure}

The ion transmission through the funnel was also tested as a function of background gas pressure. A manual valve connected to the roots pump was used to control the pressure in the chamber, ranging from 1 mbar down to 0.3 mbar. Fig. \ref{funnel-pressure} shows the measured intensity for pyrene cations (C$_{16}$H$_{10}^+$, mass~=~202~amu) as function of the pressure (black circles). This test showed that the best transmission is obtained at the lower residual gas pressure of 0.3 mbar achieved in the ion funnel with our current pumping equipment. The same conclusion applies to ions with other masses. 

Simulations (red squares in Fig. \ref{funnel-pressure}) of ion trajectories through the funnel have shown that the maximum current could be reached at a slightly lower pressure of $\approx 0.2$ mbar. These simulations use the ion funnel model in SIMION 8, adapted to our geometry and settings (RF voltage = 580 V$_{PP}$, $f = 1.75$ MHz and DC gradient $=-0.438$ V/cm). We included in the simulations the conductance limiter provided by the octupole trap entrance DC electrode following the ion funnel. Without this aperture and with the optimal pressure, 100 \% transmission can be reached, as was observed with other ions funnels \cite{Kelly2010,Lynn2000}. With the conductance limiter, our simulations indicate that about 60 \% of the ions can be transmitted into the octupole trap at the optimum pressure (see vertical axis on the right in Fig.~\ref{funnel-pressure}). A DC electrode with a bigger aperture ($>2$ mm) would improve the ion transmission through the apparatus. However, the gas flow on the funnel axis should then be reduced to avoid increases in the background gas pressure in the following sections. To do this, the relatively short section with plates of constant inner diameter in our funnel could, for example, be extended \cite{Kim2000}. A jet disrupter (disk electrode) can be placed on the funnel axis (see Kim \textit{et al.} \cite{Kim2001}) to disperse the gas flow coming from the capillary. Or an enclosure could be mounted around the source's needle and the entrance of the capillary to control the atmosphere.

\subsection{Absolute destruction cross section measurements}

Absolute destruction cross sections can be measured using the beam attenuation method. Collisions in the gas cell that lead to destruction of the molecular ions reduce the intensity of the primary beam, with the attenuation depending exponentially on the pressure $p$ in the gas cell:
\begin{equation} \label{attenuation}
\frac{I(p)}{I_0} = \exp(-\rho l \sigma)
\end{equation}
where $I(p)$ is the intensity of the primary beam (the beam of non-fragmented ions) after the gas cell when the pressure is $p$ and $I_0$ is the primary beam intensity for $p=~0$. Here $\rho= p/k_BT$ is the number density of the neutral gas where $T$ is the temperature of the gas and $k_B$ is the Boltzmann constant. The effective length of the gas cell is $l$ and $\sigma$ is the absolute total ion destruction cross section.

\begin{figure}[h!t]
\includegraphics[width=.5\textwidth, trim = 0 0 0 2cm]{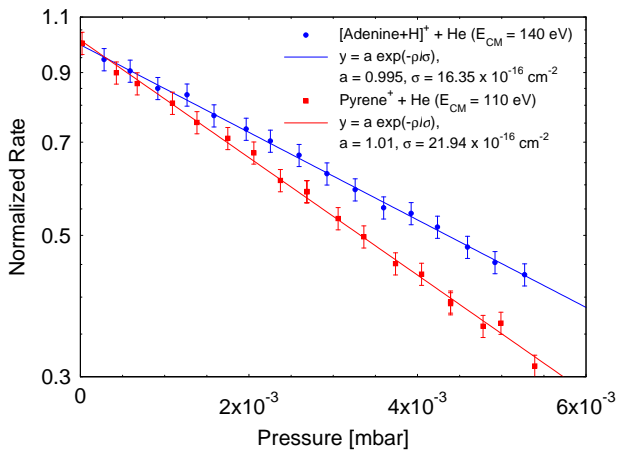} 
\caption{Attenuation measurements of pyrene cations C$_{16}$H$_{10}^+$ (m/z = 202 amu) colliding with He at 110 eV (red squares) and protonated-adenine (m/z = 136 amu) colliding with He at 140 eV (blue circles) center-of-mass collision energies. The normalized rate $I/I_0$ is on a log scale. The solid lines are exponential fits from which we extract the cross sections. The statistical errors are one standard deviation. The error bars on the pressure measurements are smaller than the symbols (less than 0.5\%).} 
\label{totXsec}
\end{figure}

In order to reduce the uncertainties due to beam fluctuations, we switch the beam between the AFEM and the gas cell with a 100 ms repetition rate so that $I$ and the AFEM count rate (which is always proportional to $I_0$) are probed a large number of times during each individual pressure setting. In this way, we are typically able to measure values of $\sigma$ with statistical uncertainties of 1\% or less. In addition, there is a systematic uncertainty connected to the uncertainty in the effective gas cell length of about 5 \% ($40 \pm 2$ mm). Fig. \ref{totXsec} shows a measurement of the beam attenuation as a function of the pressure in the gas cell for pyrene cations C$_{16}$H$_{10}^+$ (m/z = 202 amu) and protonated-adenine (m/z~=~136~amu) colliding with He at 110 eV and 140 eV center-of-mass collision energies, respectively. The corresponding destruction cross sections are proportional to the slopes of the attenuation curves in the figure. In the examples shown in Fig. \ref{totXsec}, we get a total cross section $\sigma = 21.94 \pm 1.32 \times 10^{-16}$~cm$^{-2}$ for pyrene cations and $\sigma = 16.35 \pm 0.98 \times 10^{-16}$~cm$^{-2}$ for protonated-Adenine.

\subsection{Collision Induced Dissociation mass spectra} \label{CID-spectrum}

Mass spectra of collision experiments are measured with pressures in the gas cell of about $5\times 10^{-4}$ mbar where mostly single collision processes take place. These spectra are obtained by ramping all electrostatic elements (including the lenses) in the energy analyzer. This allows for a scan of the ion kinetic energy, resulting in a spectrum of counts as a function of the analyzer voltage. For each event, the x- and y-position of the hit and the settings of the horizontal deflector voltage are recorded. The positions and analyzer voltage are then used to calculate an $E/q$ for each individual event. A histogram of these events gives us the corresponding kinetic-energy-per-charge spectrum. Using the information on the position improves the resolution of the spectrum compared to detection in single counting mode. 

\begin{figure}[h!t]
\includegraphics[width=.5\textwidth, trim = {0cm 0cm 0cm 0cm}]{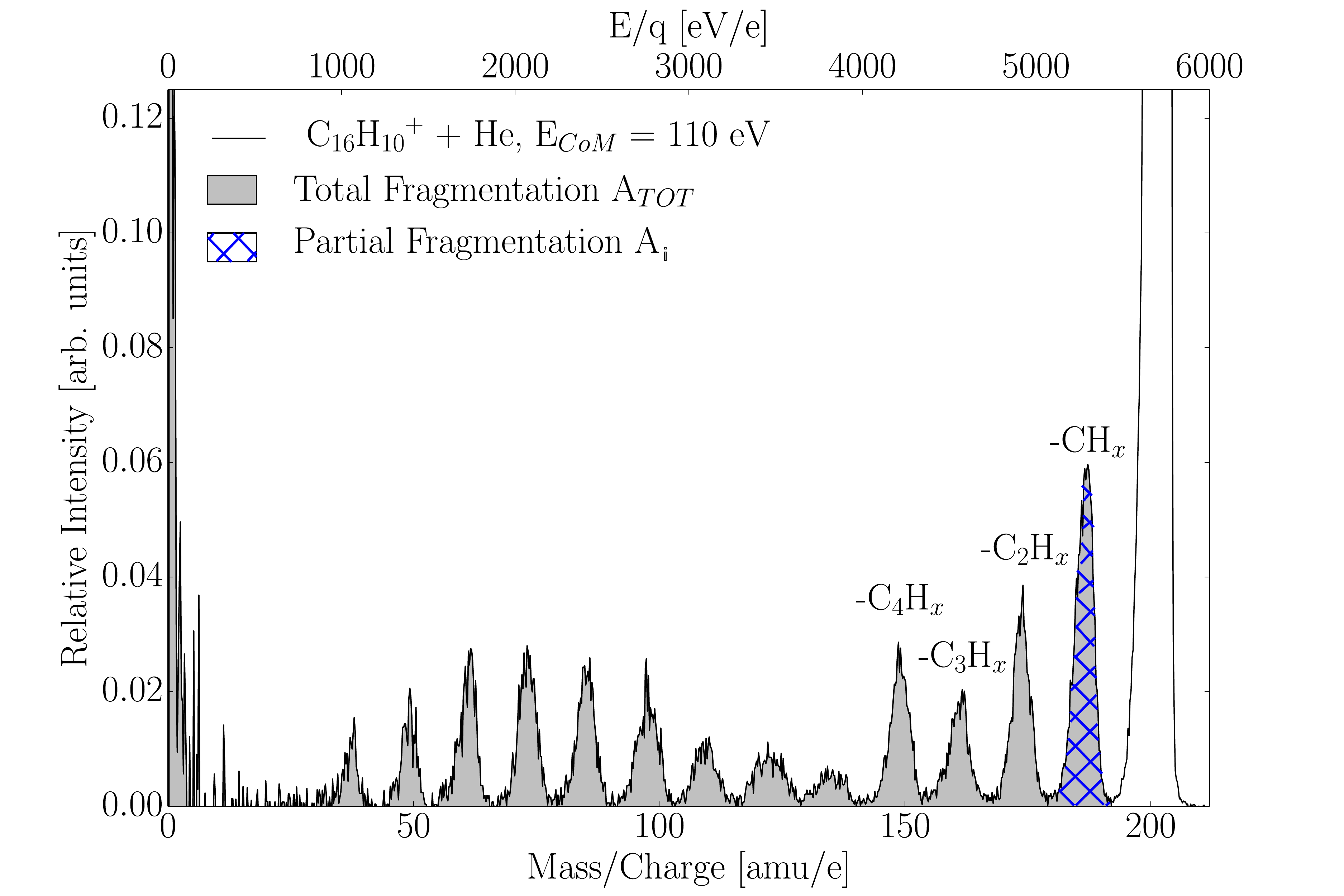} 
\caption{CID mass spectrum of pyrene cations in collision with He at a pressure of $5\times 10^{-4}$ mbar and at 110 eV in the centre-of-mass frame. The intensities are corrected by the factor $E_p/E_f$ (see text). The gray area correspond to the absolute total fragmentation cross section, while the dashed blue area gives the corresponding partial fragmentation cross section (assuming the detection efficiency is the same for all fragments).}
\label{CIDspectra-fig}
\end{figure}

As the voltages generating the deflecting electrical fields in our energy-per-charge analyzer are scanned at a constant speed ($\Delta V/\Delta t = const.$), the time it takes to sweep a beam of given energy spread across a slit is inversely proportional to its kinetic energy $E$. Therefore, the spectral intensities recorded under these conditions must be corrected with the factor $E_p/E_f$, where $E_f$ is the kinetic energy of the fragment and $E_p$ is the kinetic energy of the primary beam (intact ions). After this step and assuming that the velocity of the fragment is equal to the velocity of the primary beam, the spectrum is converted to a mass-per-charge spectrum, as shown in Fig. \ref{CIDspectra-fig}.

\subsection{Absolute partial cross sections}

Collisions between positively charged ions and He at the present low collision energies are very unlikely to lead to electron transfer from the target and thus neutralization of the the primary ion is also very unlikely. Thus, for each fragmentation event there will be one positively charged fragment and the absolute cross section $\sigma_i$ for producing a given fragment $i$ is simply 
\begin{equation}
\sigma_i = \sigma\frac{A_i}{A_{TOT}}.
\end{equation}
where $\sigma$ is the total fragmentation cross section and $A_i$ and $A_{TOT}$ are, respectively, the areas of the peak $i$ as shown in dashed blue in Fig.~\ref{CIDspectra-fig} and of all fragmentation peaks (excluding the primary ions) as indicated in gray in the example in Fig.~\ref{CIDspectra-fig}.

\begin{figure}[h!]
\includegraphics[width=.5\textwidth]{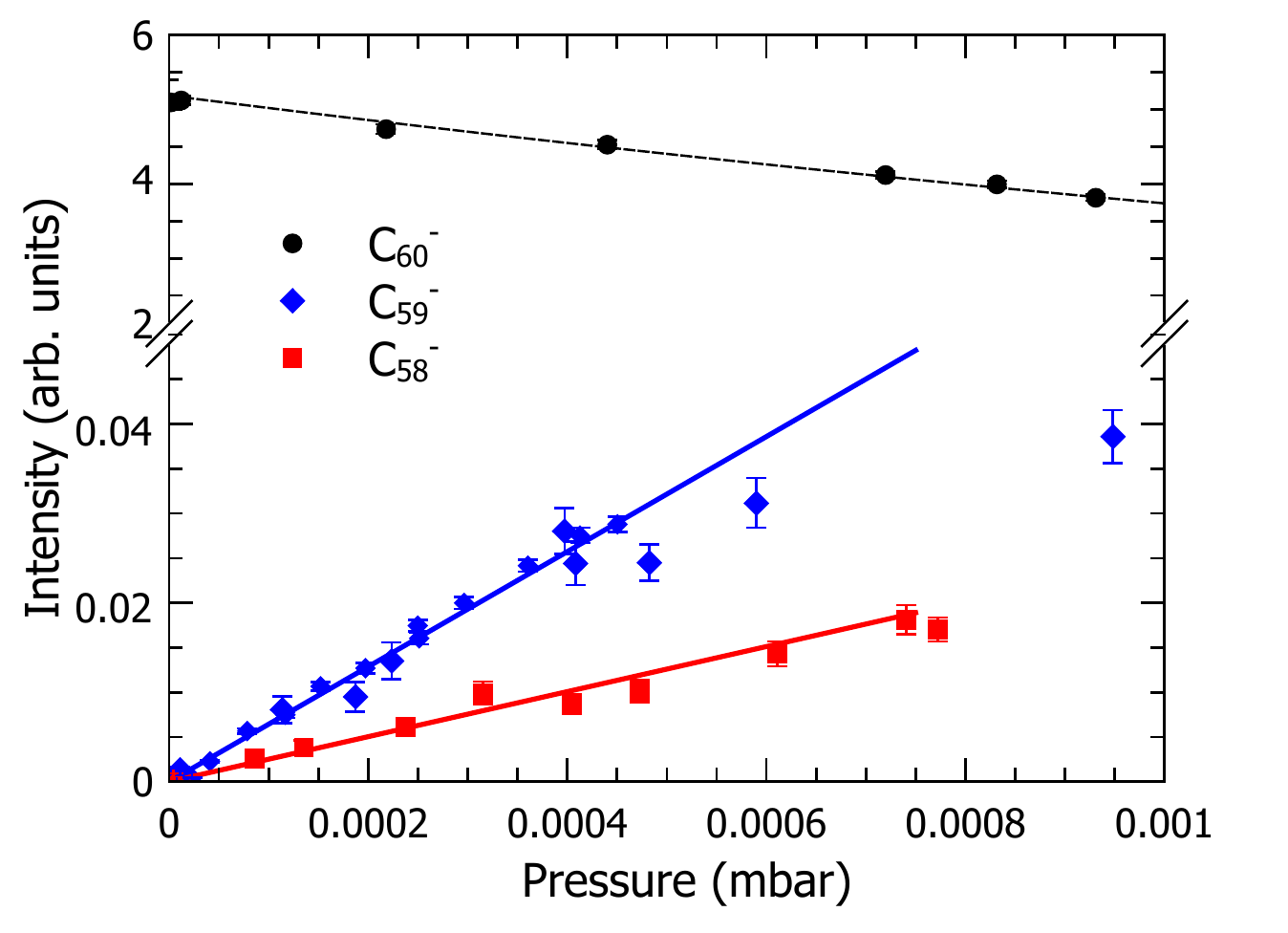}  
\caption{Attenuation measurement of 10 keV C$_{60}^-$ colliding with He and measurements of increasing intensities for two of the fragments, C$_{59}^-$ and C$_{58}^-$, at different pressures in the gas cell. Error bars are one standard deviation, the black dashed line is an exponential fit, the red and blue full lines are linear fits.}
\label{figXsec}
\end{figure}

This method can not be used with negative ion projectiles since some of the destruction mechanisms, likely to be important or even dominant (\emph{e.g.} collisional detachment creating neutrals), are in general not included in the measured fragmentation spectrum, and we therefore apply the method described below: 

At sufficiently low pressures $p$ in the cell to have only single collisions, the primary beam intensity after the cell may be written:
\begin{equation}
I(p) \simeq I_0 - I_0 \frac{\sigma l}{k_B T} p
\end{equation}
which means that the yield of any product with partial cross section $\sigma_i$ ($\sum_i \sigma_i = \sigma$) increases as 
\begin{equation}
\frac{dI_i(p)}{dp} = I_0 \frac{\sigma_i l}{k_B T}. 
\end{equation}

As for the total destruction cross section, the parent beam is alternately sent to the gas cell or to the AFEM detector to allow the normalization of the fragment intensity to the total beam intensity before fragmentation. Figure \ref{figXsec} shows three different measurements, i.e. the attenuation of C$_{60}^-$ (black circles) colliding at~10~keV with He, the corresponding increases of C$_{59}^-$ fragments (blue diamonds) and the increases of C$_{58}^-$ (red squares) fragments with increasing pressure $p$. The dashed line is an exponential fit, and the full lines are linear fits from which we extract the total and partial cross sections for these individual fragmentation channels, respectively.

\subsection{Resolution}

\begin{figure}[h!t]
\includegraphics[width=.5\textwidth]{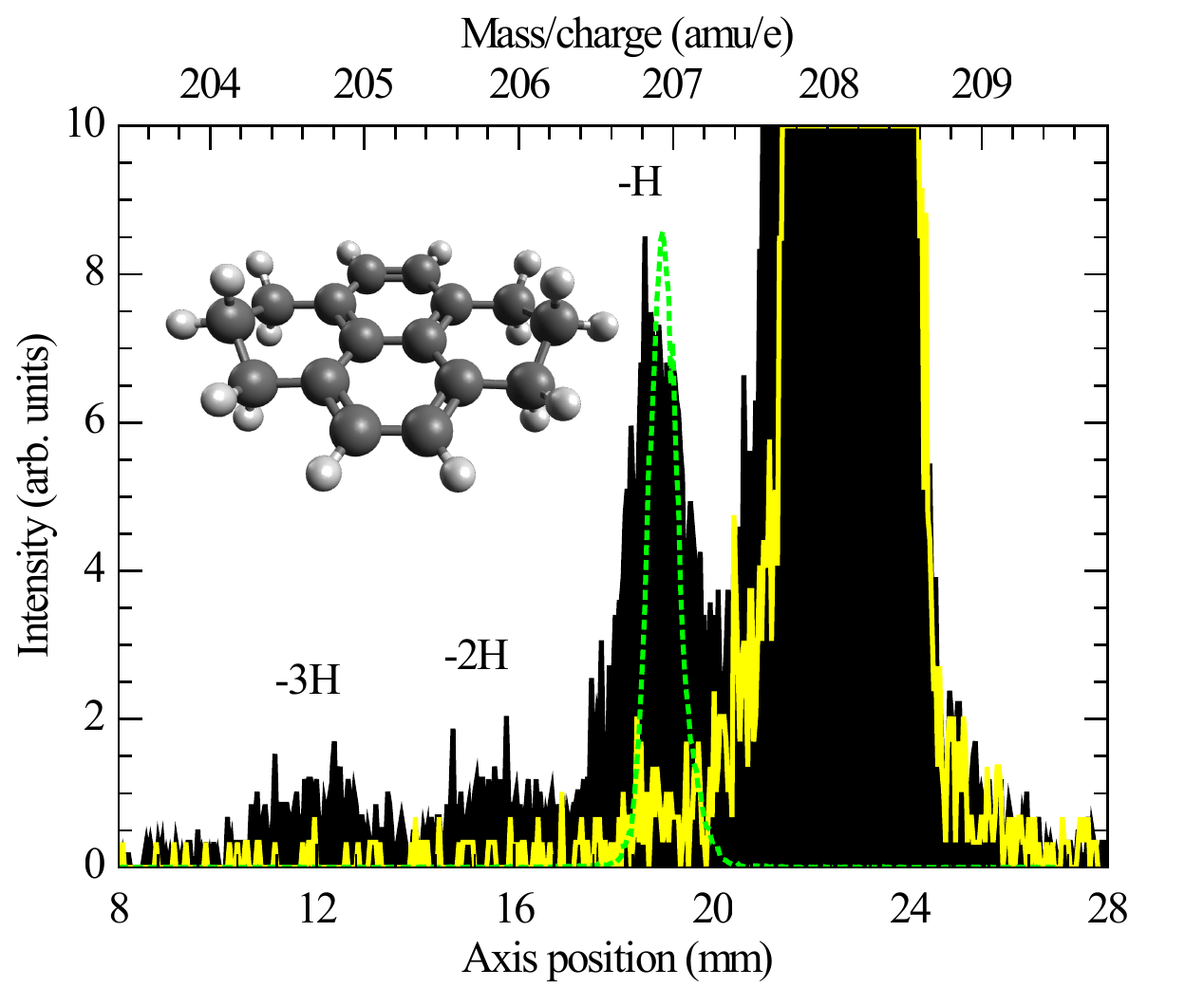}  
\caption{Count rate as function of horizontal axis position on the MCP for C$_{16}$H$_{16}^+$ at 12.00 keV without gas (yellow), with $10^{-4}$ mbar He (black) and at 11.94 keV without gas (dashed green). The energies refer to energies in the laboratory frame of reference. The cut-off peak around 208 amu is the parent hydrogenated pyrene cations C$_{16}$H$_{16}^+$ shown in the inset. The labels on top of the peaks indicate the number of hydrogen atoms lost from the parent ion. The scale on the top indicates the mass of fragments with the same velocity as the primary beam.} 
\label{figHloss}
\end{figure}

Experiments in which high mass resolution is required often suffer from limited angular acceptances making it very difficult to collect fragments emitted with large angles and thus measure the exact yield of each fragmentation processes. For this reason, the energy analyzer in this experiment has a large angular acceptance, allowing measurements of absolute cross sections. A higher mass resolution may, however, also be achieved with the present setup by closing the vertical slits in front of the analyzer (see Figure \ref{setup2}). One example of this is shown in Fig. \ref{figHloss}.

The loss of single hydrogen atoms from the parent molecular ions was observed using 1 mm wide slit and a $12.00 \pm 0.05$ keV beam of hydrogenated pyrene cations (C$_{16}$H$_{16}^+$, mass = 208 amu, see inset in Fig. \ref{figHloss}). In Figure~\ref{figHloss}, we show a histogram of the counts on the horizontal axis of the MCP detector. Measurements were done at fixed voltages of the analyzer without gas in the gas cell (in yellow) and then repeated with He at $10^{-4}$~mbar (in black). The measurements without gas (yellow curve) show the distribution of the horizontal position with the 12.00 keV primary beam. With $10^{-4}$~mbar of He in the gas cell (black position distribution), we clearly see three additional peaks corresponding to the loss of one, two, and three H-atoms from the parent ion. This was determined based on the fact that after collisions, the fragments with lower masses have less kinetic energy by amounts that are close to the ones expected for losses of 1H, 2H, and 3H when the velocities of the large fragments (C$_{16}$H$_{15}^+$, C$_{16}$H$_{14}^+$ and C$_{16}$H$_{13}^+$) are the same as for the C$_{16}$H$_{16}^+$ primary ions. Keeping the parameters on the energy analyzer (lenses and deflectors) fixed and optimized for a 12.00 keV beam, we changed the beam energy to $11.94 \pm 0.05$ keV which is the expected energy for 207 amu fragments following collisions of 12.00 keV C$_{16}$H$_{16}^+$ with He when the fragment has the same velocity as the primary beam. The peak thus obtained (dashed green) overlaps with the peak labeled -H in Fig. \ref{figHloss}, i.e. hydrogenated pyrene ions having lost one hydrogen. The slight shift and broadening of the black data in relation to the green data in Fig. \ref{figHloss}, reflects the kinetic energy loss in the collision process leading to H-loss. 

For ions having lost one or several heavy atoms like C and/or~N, the spread in energy due to the collisional process is larger than for the loss of H atoms. However, as the beam/collision energy decreases, we observe a broadening and a shift towards lower apparent masses of the peaks in the mass spectra (see Figure~1 in Stockett {\em et al.} \cite{Stockett2015}). This is due to conversion of larger fractions of the available energy into internal energies resulting in a decrease of the translational energy of the fragments. If the beam energy gets too low, the separation between the peaks become comparable to the energy spread. The broadening and a shift of the peaks at low collision energies has to be taken into account when measuring partial cross sections with the general method described for negative ions in the previous section.

\section{Summary}

An ion source platform for the production and handling of large, fragile molecular ions using an electrospray ion source has been developed at the DESIREE infrastructure at Stockholm university. On the platform, an ion funnel, ion-beam guides and traps, a quadrupole mass filter and a four-way quadrupole deflector are used to form continuous or pulsed beams of mass selected ions. The voltages on the four-way deflector can be switched to send the beam either in a detector for ion-beam intensity measurements, a trap for ion cooling, or the experimental setups for collision studies. The CID setup is designed for large angular acceptance experiments. We have provided detailed descriptions of the various components of these setups and we have described results from critical tests. 

The setup has, in different specific configurations, been used to study fragmentations of complex molecules and a range of new findings have already been reported. Some of these have lead to important new insights concerning molecular fragmentation mechanisms. Studies of fragmentations and energy-transfer mechanisms in native and hydrogenated PAHs \cite{Chen2014, Stockett2014a, Stockett2014b, Stockett2015, Stockett2015b, Gatchell2015, Wolf2016}, fullerenes \cite{Gatchell2014,Stockett2018} and biomolecules \cite{Giacomozzi2016,Kostya2015,Kostya} have been performed with the EIS-Lab setup for center-of-mass energies between 10 eV and 1 keV. In this region, nuclear stopping processes typically dominate the energy transfer to the molecule similar to what is believed to occur in supernova shockwaves \cite{Micelotta2010,Tappe2012}. This laboratory work has revealed the importance of single carbon knockout in PAHs \cite{Stockett2014a, Stockett2014b, Stockett2015} and atom capture by fullerenes \cite{Stockett2018}. Other experiments have demonstrated the importance of knockout process in molecular systems with low dissociation energies such as hydrogenated pyrene ions \cite{Gatchell2015,Wolf2016} and tetraphenylporphyrin ions \cite{Giacomozzi2016}. Such non-statistical dissociation processes lead to highly reactive fragments that may later react and form larger species. Similar processes are possible pathways to DNA damage caused by radiation. 

This versatile setup can be used in continuous or pulsed mode. The expected short ion bunches ($\approx$ 15 {\textmu}s pulse length) created by the combination of an octupole trap with ejection electrodes and a cryogenic RF ring-electrode trap with He buffer gas could be injected in ion-beam storage devices for experiments with cold stored complex ions. We plan to test the pulsed-mode operation and then use the ion source platform (the whole assembly shown in Fig.~\ref{setup1}) as an injector for DESIREE \cite{Thomas2011,Schmidt2013} and in an electrostatic ion trap \cite{Schmidt2001} mounted between the electrostatic deflectors and the detector in the CID beamline.

\section*{Supplementary material}

See supplementary material for a list of the commercial components on the high voltage platform and beamline. This includes the vacuum equipment. A description of the Extrel CMS quadrupole mass filter is also available.

\section*{Acknowledgments}

This work was performed at the Swedish National Infrastructure, DESIREE (Swedish Research Council Contract No. 2017-00621). It was further supported by the Swedish Research Council (grant numbers 2014-4501, 2015-04990, 2016-03675, 2016-04181, 2016-06625).

\section*{References}

\bibliographystyle{apsrev4-1}
\bibliography{eislab}{}

\end{document}